# IP Traceback for Flooding attacks on Internet Threat Monitors (ITM ) Using Honeypots

K.Munivara Prasad[1] and A.Rama Mohan Reddy[2]   V Jyothsna[3]

[1]Department of Computer Science and Engineering, Rayalaseema University, Kurnool
`prasadkmv27@gmail.com`
[2]Professor and Head,Departmentof Computer Science and Engineering,SVUCE,SV University, Tirupati
`ramamohansvu@yahoo.com`
[3]Department of Information technology, Sree vidyanikethan Engg.College,Tirupati
`jyothsn1684@gmail.com`

## ABSTRACT

*The Internet Threat Monitoring (ITM) is an efficient monitoring system used globally to measure, detect, characterize and track threats such as denial of service (DoS) and distributed Denial of Service (DDoS) attacks and worms. . To block the monitoring system in the internet the attackers are targeted the ITM system. In this paper we address the flooding attack of DDoS against ITM monitors to exhaust the network resources, such as bandwidth, computing power, or operating system data structures by sending the malicious traffic. We propose an information-theoretic frame work that models the flooding attacks using Botnet on ITM. One possible way to counter DDoS attacks is to trace the attack sources and punish the perpetrators. we propose a novel traceback method for DDoS using Honeypots. IP tracing through honeypot is a single packet tracing method and is more efficient than commonly used packet marking techniques.*

## KEYWORDS

*Internet Threat Monitors (ITM), DDoS, Flooding attack,IpTrcing, Botnet and Honeypot.*

## 1. INTRODUCTION

The Internet was initially designed for openness and scalability. The infrastructure is certainly working as envisioned by that yardstick. However, the price of this success has been poor security. On the Internet, anyone can send any packet to anyone without being authenticated, while the receiver has to process any packet that arrives to a provided service. The lack of authentication means that attackers can create a fake identity, and send malicious traffic with impunity. A denial-of-service (DoS) attack [2] is an explicit attempt by attackers to prevent an information service's legitimate users from using that service. These attacks, attempt to exhaust the victim's resources, such as network bandwidth, computing power, or operating system data structures. Flood attack, Ping of Death attack, SYN attack, Teardrop attack, DDoS, and Smurf attack are the most common types of DoS attacks. The hackers who launch DDoS attacks typically target sites or services provided by high-profile organizations, such as government agencies, banks, credit-card payment gateways, and even root name servers.

A flooding-based Distributed Denial of Service (DDoS) attack is a very common way to attack a victim machine by sending a large amount of unwanted traffic. Network level congestion control can throttle peak traffic to protect the network. Network monitors are used to monitor the traffic in the networks to classify them as genuine or attack traffic and also these monitors gives the traffic as an input to several DDoS detection algorithms for detection of DDoS attacks. However, it cannot stop the quality of service (QoS) for legitimate traffic from going down because of attacks. Two features of DDoS attacks hinder the advancement of defence techniques. First, it is hard to distinguish between DDoS attack traffic and normal traffic. There is a lack of an effective differentiation mechanism that results in minimal collateral damage for legitimate traffic. Second, the sources of DDoS attacks are also difficult to find in a distributed environment. Therefore, it is difficult to stop a DDoS attack effectively.

Threat monitors are used to monitor the packet flow at the routers level or the network level to detect the malicious traffic in the network to avoid the security threats. The ITM is a distributed, globally scoped, the goal of the ITM is to measure, characterize, and track a broad range of Internet threats. The fundamentally distributed nature of this architecture allows the ITM to monitor diverse addresses and topologies. The design of Internet threat monitors (ITM) has the major effort in the internet to prevent the users from the security attacks. The ITM has two major components data center and the monitors. Each monitor covers the range of ip addresses and records the traffic and sends the traffic logs periodically to the data center. The data center collects the traffic logs and publishes the reports to ITM system users.

The monitor locations of an ITM system can be compromised by introducing several attacks by the attackers which includes Localization attacks [1] and DDoS Attacks[3][6] .Which exploits some vulnerability or implementation bug in the software implementation of a service to bring that down or that use up all the available resources at the target machine or that consume all the bandwidth available to the victim machine, this is called as Bandwidth attacks. It is clear that DDoS attacks will not stop or scale down until they are properly addressed.

One possible way to counter DDoS attacks is to trace the attack sources [17] and punish the perpetrators. However, current Internet design makes such tracing difficult in two aspects. First, there is no field in the IP header that indicates its source except for the IP address, which can be easily spoofed by an attacker. Second, the Internet is stateless in that it does not keep track of the path traversed by a packet. Recently, efforts are made to change one or both aspects to allow for tracing packets to their sources, known as IP Trace- back.

IP tracing methods can be classified [18] into two categories preventive and reactive. Precautionary steps can be taken by the preventive method for DoS and DDoS attack prevention. The goal of the reactive methods is to identify the source of attacks, for that these method provides the wide range of solutions. The reactive methods are more efficient to identify the source even the attacker spoofs their addresses.

In this paper we introduce an information theoretic frame work to model existing flooding attacks in ITM system monitors. In the flooding attack the attacker sends the large volume of unwanted traffic to the targeted monitor for this he uses the botnet. Based on the Information-theoretic model we propose an effective IP tracing approach to trace the attack source using Honey pots.

## 2. RELATED WORK

Probing traffic based Localization attack [7][8] in which an attacker sends high rate short length port scan messages to the targeted network to compromise the monitor locations in ITM system. Then, attacker queries the data center to determine whether a short spike of high-rate traffic appears in the queried time-series data, for confirmation of the attack.

A steganographic localization attack [9] an attacker launches a stream of low-rate port-scan probing traffic which is marginally modulated by a secret Pseudonoise (PN) code. While the low-rate property prevents the exhibition of obvious regularity of the published traffic data at the data center, based on the carefully synchronized PN code, the attacker can still accurately identify the PN-code-modulated traffic in the retrieved published traffic data from the data center. Thereby, the existence of monitors in the targeted network can be compromised. To this end, the PN-code-based steganographic attack presented in our paper can be understood as a covert channel problem [10], because the attack traffic encoded by a signal blends into the background traffic and is only recognizable by the attacker which knows the secret pattern of the PN code.

In [1] introduced the information theoretic framework to evaluate the effectiveness of the localization attacks by using the minimum time length required by an attacker to achieve a predefined detection rate as the metric. But this frame work is defined in specific to the localization attacks only; they are not given any solution for other DDoS attacks. The frame work allows the ITMs which are registered within the data center given, and the access is restricted to that private region only. But public access of the ITMs and data center allows more scope to provide security against different attacks.

Both the IP backtracking [18] methods preventive and reactive have their own drawbacks while tracing the source of an attacker. Preventive methods can provide only the precautionary steps and these do not support scalability and have low compatibility and high router overhead. In reactive methods PPM, DPM, ICMP [19][22][23]Trace back and Hash based methods do not provide better scalability and some of them require more memory at routers, which is not practically possible.

## 3. PROPOSED WORK

In [1] the authors define a model in which the ITMs in the networks sends the traffic logs periodically to the data center and the data center collects the traffic logs and publishes the reports to ITM system users which are registered, that means it creates the private environment or region .In the private region the scope for DDoS attacks are very less, and they are restricted this model only for Localization attacks. In this section we have defined a model which will provide the following extensions.

**Public accessing:** Public accessing of the data center increases the network usage and provides better communication with the outside world rather than private environment. In this any user from outside the private region can get the communication with the private network, if the user is genuine he can get the status of the monitor before sending the data to internal monitors, to avoid the attacks. If the user is an attacker, then this status information can be misused to perform the attacks on the monitor. The data center sends the status information to any users (public or private) based on the request query, but the private (internal) users can get the highest priority.

**Usage of Botnets for Flooding Attack:**

A denial-of-service (DoS) attack is an explicit attempt by attackers to prevent an information service's legitimate users from using that service. In a DDoS attack, these attempts come from a large number of distributed hosts that coordinate to flood the victim with an abundance of attack packets simultaneously. The attacker may use the botnets [11], [12] and other alternatives to launch the attack.

### 3.1 Flooding:

*Launching a flooding attack:* Once the DDoS network has been set up and the infrastructure for communication between the agents and the handlers established, all that an attacker needs to do is to issue commands to the agents to start sending packets to the victim host. The agents try to send unusual data packets (all TCP flags set, repeated TCP SYN packets, Large ICMP packets) to maximize the possibility of causing disruption at the victim and the intermediate nodes. There are certain basic packet attack types which are favorites of the attack tool designers. All the attack tools use a combination of these packet attack types to launch a DDoS attack. The basic attack types are

*i) TCP floods:* A stream of packets with various flags (SYN,RST, ACK) are sent to the victim machine. The TCP SYN flood works by exhausting the TCP connection queue of the host and thus denying legitimate connection requests. TCP ACK floods can cause disruption at the nodes corresponding to the host addresses of the floods as well. Also the one known tool that uses TCP ACK flooding (mstream [13]) has been known to cause disruptions in a router even with a moderate packet rate. Both TCP SYN flooding and the mstream attack constitute a group of attacks known as asymmetric attacks (Attacks where a less powerful system can render a much more powerful system useless).

*ii) ICMP floods (e.g ping floods):* A stream of ICMP packets is sent to the victim host. A variant of the ICMP floods is the Smurf attack in which a spoofed IP packet consisting of an ICMP ECHO_REQUEST is sent to a directed broadcast address. The RFC for ICMP specifies that no ECHO_REPLY packets should be generated for broadcast addresses, but unfortunately many operating systems and router vendors have failed to incorporate this into their implementations. As a result, the victim host (in this case the machine whose IP address was spoofed by the attacker) receives ICMP ECHO_REPLY packets from all the hosts on the network and can easily crash under such loads. Such networks are known as amplifier networks and thousands of such networks have been documented.

*iii) UDP floods:* A huge amount of UDP packets are sent to the victim host. Trinoo is a popular DDoS tool that uses UDP floods as one of its attack payloads.

### 3.2 BOTs

Studying the evolution of bots and botnets provides insight into their current capabilities. One of the original uses of computer bots was to assist in Internet Relay Chat (IRC) channel management [14]. IRC is a chat system that provides one-to-one and one-to-many instant messaging over the Internet. Users can join a named channel on an IRC network and communicate with groups of other users. Administering busy chat channels can be time consuming, and so channel operators created bots to help manage the operation of popular channels. One of the first bots was Eggdrop, which was written in 1993 to assist channel operators [1].

In time, IRC bots with more nefarious purposes emerged. The goal of these bots was to attack other IRC users and IRC servers. These attacks often involved flooding the target with packets (i.e., DoS attacks). The use of bots helped to hide the attacker because the attack packets were sent from the bot rather than directly from the attacker (assuming a non-spoofed attack). This new level of indirection also allowed multiple computers to be grouped together to perform distributed attacks (DDoS) and bring down bigger targets. Larger targets required more bots, and so attackers looked for methods to recruit new members. Since very few users would agree to have their computers utilized for conducting packet floods, attackers used trojaned files and other surreptitious methods to infect other computers.

## 3.3 Botnet:

One of the biggest threats to the Internet is the presence of large pools of compromised computers, also known as botnets, or zombie (drone) armies, sitting in homes, schools, businesses and governments around the world. Under the control of a single (or a small group of) hacker, commonly known as a botmaster, botnets are often used to conduct various attacks, ranging from Distributed Denial-of-Service (DDoS) attacks to e-mail spamming, keylogging, click fraud, and spreading new malware. Unlike other types of attacks, botnets which may be comprised of thousands of compromised hosts can assemble a tremendous amount of aggregate computing power and can perform a variety of attacks against a wide range of targets. For instance, a botmaster can command each zombie participant in a botnet to launch spamming emails, perform some sort of credit-card theft (gleaned from surreptitiously planted keyloggers), and launch DDoS attacks simultaneously against thousands of computer hosts. Because of this, hackers are increasingly interested in using botnets to launch attacks to maximize their financial gains. At the same time, the degree of destruction caused by hackers using botnet attacks is hundreds of times larger than traditional, discrete attacks. Since the appearance of botnets, new and sophisticated software modules are added into existing botnet tools every day, offering a variety of ways to compromise computers and launch potentially much more harmful attacks.

Recently, the threats presented by botnets are just beginning to be realized. The Internet community at-large, law enforcement organizations, individual users, and enterprises alike are all beginning to discuss methods to defeat botnets, perhaps the single biggest security threat to today's Internet community.

**Attacking Behavior :** In the course of an attack, botnets normally generate a large amount of abnormal traffic, which in turn can facilitate easy detection. Furthermore, if more effort is spent on understanding the attacking behaviors, a lot more information can reveal important intelligence, including the nature of a botnet, the purpose of the hackers, and even the origins of the hackers. Based on this information, we can propose more effective countermeasures (e.g., detection, prevention and remedy plans). In this paper, we discuss attacking behaviors from the following four aspects:

- Infecting new hosts
- Stealing personal information
- Phishing and spam proxy
- DDoS

**Infecting new hosts:** Botnets often recruit new hosts using similar approaches as those for other malware (i.e., virus and worm). One of the methods that botnets use to compromise new hosts is through social engineering and distribution of malicious emails. In a common scenario, a botnet may distribute email messages with malware attached, or perhaps an embedded link to a malware binary located elsewhere. Social engineering techniques are used to trick computer users into executing the malware, which leads to the compromise of hosts.

**Stealing Sensitive Information:** Recent botnets have employed sophisticated tools to steal sensitive user information from compromised hosts. The most commonly used tools for stealing sensitive information are keyloggers and network traffic sniffers. Keyloggers modify host operating systems to spy on user activities and capture user key strikes**.** Network traffic sniffers monitor network traffic sent over the subnet of the compromised host. The sensitive data is logged by these tools and then compiled into digested formats. Periodically, the data will be sent to their botmasters using various communication channels. Some commonly used methods are to send data through a designated IRC channel created by a botnet and in emails to a designated email address. BKDR_WAR.B steals keystrokes on a compromised computer in this way.

**Sending Spam:** Botnets are widely used to disseminate spam for different attack purposes. Two major advantages for hackers to use botnets to distribute spam (as opposed to sourcing it from a single compromised host) are that the victims cannot trace the spam back to the source for legal action, and botnets can distribute a much larger volume of spam because of the aggregate computing power and vast availability of bandwidth. While some spam is used to distribute exploits (malware) as described in a previous subsection, some spam tricks users into visiting certain malicious websites, which install malware on their computers by exploiting Internet browser vulnerabilities.

**Distributed Denial of Service:** A DDoS attack [19] is probably one of the oldest botnet attack mechanisms. In the infancy of botnets, hackers began using botnets to launch DDoS attacks against a number of large organizations to consume all of their available platform CPU cycles and available bandwidth, effectively slowing their services down to a crawl, or knocking out their services altogether. For example, both Yahoo! and Microsoft were victimized by DDoS attacks launched by botnets in the past years. DDoS attacks still occur, but in a lesser frequency and volume. DDoS attacks have even recently been used for extortion. Botnets usually integrate a large variety attacking tools (e.g., UDP flooding, TCP SYN flooding, HTTP flooding). Some bots, such as PhatBot , even have very customized DDoS tools integrated into their code. AgoBot, SDBot, PhatBot, and many other botnets are all capable of launching DDoS attacks against a variety of targets.

**Botnet Life Cycle:** The success of any process mainly lies in how well the sequence of steps is organized. The major reason of dramatic success and spread of botnets is their well organized and planned formation, generation and propagation. The lifecycle of a botnet from its birth to disastrous spread undergoes the following phases:

**1.** Bot-herder configures initial bot parameters.
**2.** Registers a DDNS.
**3.** Register a static IP.
**4.** Bot-herder starts infecting victim machines either directly through network or indirectly through user interaction.
**5.** Bots spread.
**6.** Bot joins the Botnet through C&C server.
**7.** Bots are used for some activity (DDoS, Identity Theft etc.)
**8.** Bots are updated through their Botoperator which issues update commands.

### 3.4 IRC-based Command and Control

A bot must communicate with a controller to receive commands or send back information. One method for establishing a communication channel is to connect directly to the controller. The problem is that this connection could compromise the controller's location. Instead, the bot controller can use a proxy such as public message drop point (e.g., a well-known message board). However, because websites and other drop points can introduce significant communication latency, a more active approach is desirable. A well-known public exchange point that enables virtually instant communication is IRC.

IRC provides a common protocol that is widely deployed across the Internet and has simple text-based command syntax. There is also a large number of existing IRC networks that can be used as public exchange points. In addition, most IRC networks lack any strong authentication, and a number of tools to provide anonymity on IRC networks are available. Thus, IRC provides a simple, low-latency, widely available, and anonymous command and control channel for botnet communication. An IRC network is composed of one or more IRC servers as depicted in Figure 1.

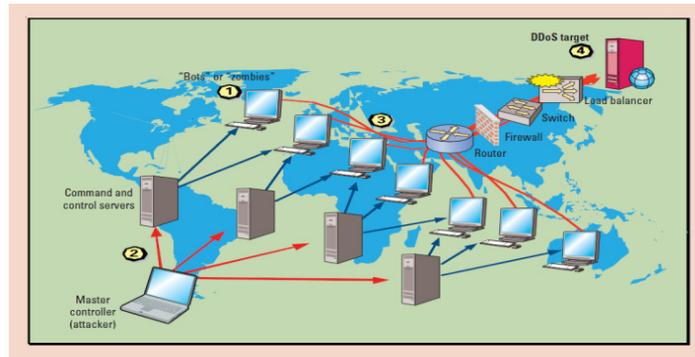

Figure 1: Compromised computers. In a distributed denial-of-service attack (DDoS), these computers serve three major roles: master controller, command and control server, and bot.

In a typical botnet, each bot connects to a public IRC network or a hidden IRC server on another compromised system. The bot then enters a named channel and can receive commands directly from a controller or even from sequences encoded into the title of the channel.

### 3.5 IP Traceback

IP tracing methods can be classified [18] into two categories preventive and reactive. Precautionary steps can be taken by the preventive method for DoS and DDoS attack prevention. The goal of the reactive methods is to identify the source of attacks, for that these method provides the wide range of solutions. The reactive methods are more efficient to identify the source even the attacker spoofs their addresses.

Ingress filtering is an example for preventive methods [19] in this the router blocks the packet if that arrives with illegitimate addresses. This method is efficient and suitable when the load on the routers is very less. In this ingress filtering method the router must have the knowledge to distinguish between legitimate and illegitimate addresses, it is difficult to provide for the routers, and forge addresses from attackers are also very difficult to identify.

Internet Control Message protocol (ICMP) [23 ] [24]is a reactive method where every router it prepare trace back message iTrace and that is directed to the same destination as the selected packet. The iTrace message itself consists of timestamp, next and previous hop information to track the attack source. ICMP Traceback is well suitable for the DoS and DDoS attacks to trace the attack source and the router is overburden to mark its address because of that router is memory less device.

Logging approach logs packets at key routers and then using data mining techniques to determine the path that the packets traversed. The advantage of this method is that it can trace an attack long after the attack has completed. The drawback of this is it requires the database support and potentially enormous resource requirements. Packet marking algorithms are used to trace the source address of the attacker by storing the packet in the routers. In Deterministic packet marking (DPM) scheme [21] each router marks all the packets passing through it with its unique identifier. Router is memory less device in this router requires more memory to store all the packets which are passed through it, practically it is very difficult. In probabilistic packet marking (PPM) scheme [20] packets are selected to mark according to the probability and stores the packets in router in which they passed it.

In Hash Based IP Traceback [22] approach the routers stores the hash value instead of storing the complete packet. The memory is minimized compared to the PPM and DPM and effectively used for DDoS attacks. But still some memory requirement is there for the routers to store the hash value of the packets, this is again practically not possible for larger flow size.

## 4. PROPOSED MODEL

In this paper we divided the entire model into two regions namely private region and public region. The Internet Threat Monitors (ITM) are distributed across the Internet and each monitor records the traffic addressed to range of IP addresses and send the traffic logs periodically to the data center. The data center then analyzes the traffic logs collected from the monitors and publishes the reports to ITM system users. The collection of monitors under the data center forms the private region because the ITMs are registered before sending the logs to the data center. Any user can get the reports of the requested ITM by sending the query request to the data center and the data center is answerable to all the ITMs which are registered.

The public region of our model specifies the unregistered users of the data center who does not have any permission to access the data center, but they can get the traffic reports related to any ITM by sending the query request to the data center. The data center scope is extended to the public domain but it can only give the traffic reports to the public users. Allowing the public users or network accessing to the data center and monitors, causes decrease in the performance because of the overload of the data center. These can be balanced by introducing the priorities to the users; the internal or private region users have the highest priority than the public users .This priorities does not disturb the existing scenario but this can enhance the service to the public domain ,this will not be a over burden to the data center..

In This section we are constructing the botnet as the public user network without having any registration with data center and performing the flooding attack on the ITM which is local to the data center.

**Generation of flooding attack with Botnet:**

A DDoS (Flooding) attack mechanism typically includes a network of several compromised computers [15]. These compromised computers serve three major role -master controller, command and control (C&C) server, and bot. An attacker prepares a DDoS attack by exploiting vulnerabilities in one computer system and making it the DDoS "master controller." From here, the attacker identifies and communicates with other compromised systems. A C&C server is a compromised host with a special program running on it, this server distributes instructions from the attacker to the rest of the bots, which form a botnet[11]. (A bot is a compromised host that runs a special program.) Each C&C server is capable of controlling multiple bots, each of which is responsible for generating a stream of packets to the intended victim. Often, the bots employed to send the flood of requests are infected with a virus that lets attackers use them anonymously.

A Flooding attack happens in several phases:

• *Discover vulnerable hosts*. To launch a DDoS attack, attackers first build a network of computers that they can use to produce the volume of traffic needed to deny services to legitimate users. To create this network, they first scan and identify vulnerable sites or hosts. Vulnerable hosts are usually those that run either no antivirus software or an out-of-date version, or those that aren't properly patched. Attackers use these compromised hosts for further scanning and compromises.

• *Establish a botnet*. After gaining access, attacker must then install attack tools on the compromised hosts to form a botnet.

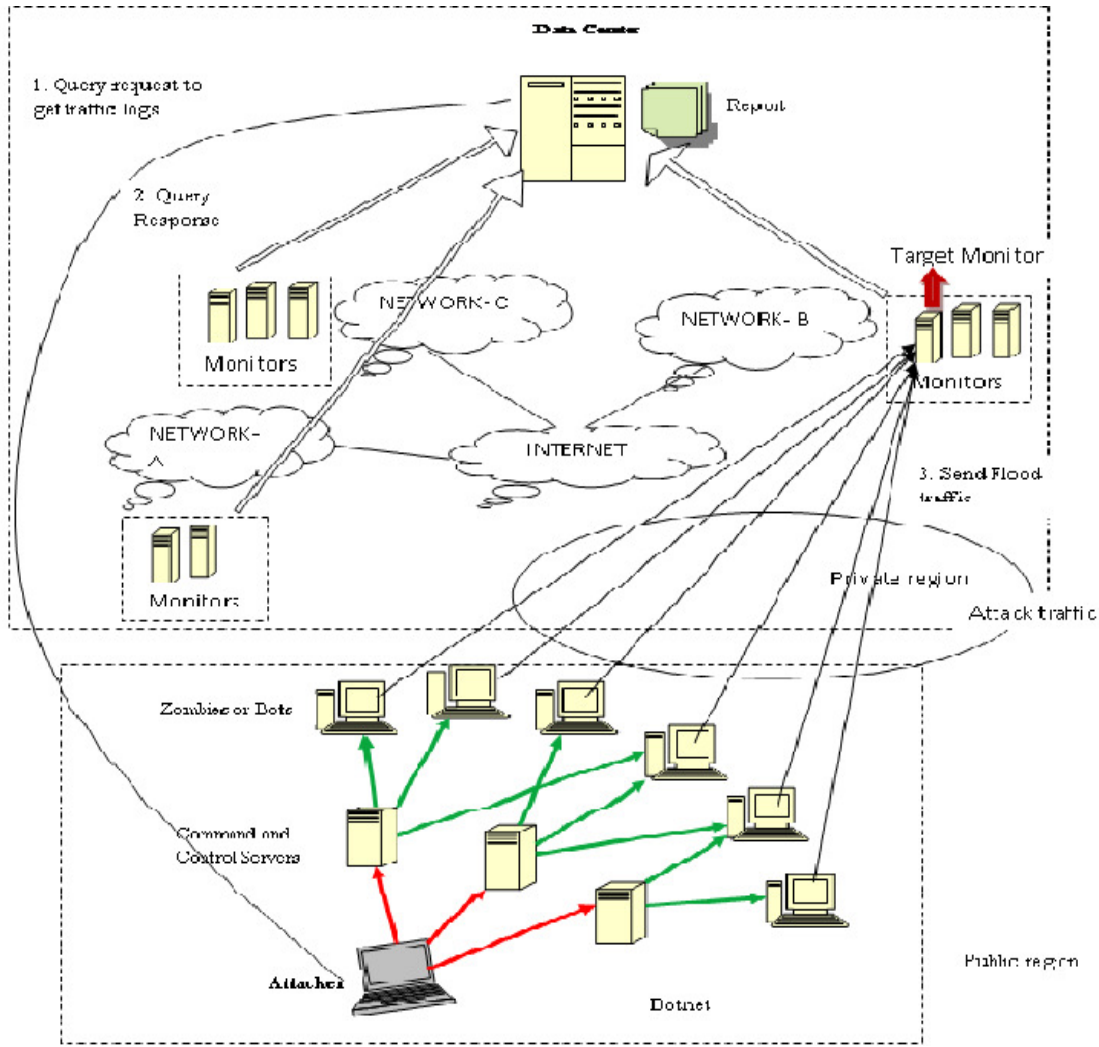

Figure 2: Work flow of flooding attacks using botnet.

• *Launch an attack*. In the next phase, attackers send commands to C&C servers for their bots to attack by sending hundreds of thousands of requests to the target simultaneously.

• *Flood a target*. In the final phase, monitor receives a flood of requests to the point where they can't operate effectively.

## 5. PREVENTION

Preventive mechanisms attempt either to reduce the possibility of DDoS attacks or enable potential victims to endure the attack without denying services to legitimate users.

• *System security mechanisms* increase a host's overall security posture and prevent it from becoming part of a botnet or a DDoS victim. Examples of system security mechanisms include reliable firewall filtering, proper system configuration, effective vulnerability management, timely patch installation, robust antivirus programs, controlled and monitored system access, and solid instruction detection.

• *Resource multiplication mechanisms* provide an abundance of resources to counter DDoS threats, such as increasing the capacity of network bandwidth, routers, firewalls, and servers. Additional examples include deploying information services at diverse network locations and establishing clusters of servers with load-balancing capabilities. Resource multiplication essentially raises the bar on how many bots must participate in an attack to be effective. While not providing perfect protection, this last approach has often proved sufficient for small- to mid-range DDoS attacks.

## Preventing Flooding Attacks

In this section we introduce a general methodology to prevent flooding attacks. It is based on the following line of reasoning:

1. To mount a successful Flooding attack, a large number of compromised machines are necessary.

2. To coordinate a large number of machines, the attacker needs a remote control mechanism.

3. If the remote control mechanism is disabled, the Flooding attack is prevented.

Our methodology to mitigate flooding attacks aims at manipulating the root-cause of the attacks, i.e., influencing the remote control network. Our approach is based on three steps:

1. Infiltrating the remote control network.

2. Analyzing the network in detail.

3. Shutting down the remote control network.

In the first step, we have to find a way to smuggle an agent into the control network. In this context, the term agent describes a general procedure to mask as a valid member of the control network. This agent must thus be customized to the type of network we want to plant it in. The level of adaptation to a real member of the network depends on the target we want to infiltrate. For instance, to infiltrate a botnet we would try to simulate a valid bot, maybe even emulating some bot commands.

Once we are able to sneak an agent into the remote control network, it enables us to perform the second step, i.e., to observe the network in detail. So we can start to monitor all activity and analyze all information we have collected.

In the last step, we use the collected information to shut down the remote control network. Once this is done, we have deprived the attacker's control over the other machines and thus efficiently stopped the threat of a flooding attack with this network. Again, the particular way in which the network is shut down depends on the type of network.

## 6. DETECTION OF FLOODING ATTACKS

In this section we present efficient way of detecting the attacks on the ITMs in the given information theoretic frame work. We divide the attack detection process into two phases, Firstly the primary detection of DDoS attacks on the ITMs and the later is the detection of flooding attacks on the ITMs.

In the primary detection phases the system detects the attacks based on traffic information aggregated from all monitors in the ITM system. If the overall traffic rate (e.g., volume in a given time interval) exceeds a predetermined threshold, the defender issues an alarm. The threshold value can be maintained either at data center or the individual ITMs based on the type of schemes used [1] in the network. In the primary detection phase the system detects some attack was happened in the network. If the detection scheme is centralized, then whenever the aggregate traffic exceeds the threshold maintained at the data center then the data center finds the attack and that attacked monitor can be identified by verifying the individual traffic logs of each ITM from the report. Otherwise if the detection strategy is distributed then each monitor maintained an individual threshold and checked the aggregate traffic regularly. If the traffic exceeds the threshold then it find the attack was happened and sends the status as attacked to the data center. After getting the attacked status from the ITM the data center blocks the corresponding ITM and displays the status of the ITM as blocked in the status reports, which will avoids the further traffic to or from the attacked ITM with the rest of the networks.

The second stage of detection specifies the detection of the flooding attacks. Once the attack is conformed then the data center identifies the attacked monitor and the traffic logs will be handover to the flooding detection phase. In this paper the flooding attacks are generated using botnet, so botnet tracking is required to detect and block the flooding attacks on the attacked ITM.

In this section we define the approaches for detecting the botnet. Once the botnet is successfully identified and blocked then automatically the flooding attacks can be avoided. In this connection the honeypots play the major role to block the botnet by identifying the command and control through the IRC server.

### 6.1 .IP Traceback Using Honeypot
IP Tracing scheme can be divided into two main parts: [25] The Honeypot subsystem and The Attack Tracing system.

The honeypot subsystem part includes honeypots and log reports with data center. In this part the deployment of honeypots, the inducing and entrapping process, and the traceback initiate request are accomplished. Attack Tracing System includes ITM(s), this part is in charge of the traceback data collection, and statistical inquiry and route rebuild computation.

### 6.1.1 The Honeypot Subsystem

The honeypot subsystem consists of 4 functional modules: the network deceive module, the information capture module, the information control module and the communication control module.

The network deceive module deals with the deceits and inducement against incoming traffic flow, in any type of security threats, opening ports and setting sensitive information. The information capture module monitors and records all activities in the honeypot. Once the status of ITM in data center reports set to "blocked", it will post a traceback request to the communication control module. The traceback request would contain such information: the ITM name, the flow size of the attack, the visit time and flow destination IP address, and so on.

The information control module can be used to restrict the honeypot contact activity. Once the attack is detected, it guarantees other hazard would be brought in. In practice, such function can be accomplished by honeypot bandwidth restriction and routers-firewall cooperation. The communication control module transfers the traceback request to the trace service console and vise versa, and then receives the traceback result from the service console.

### 6.1.2 Attack Tracing System

This is the combination of Trace service console and data center. This part is in charge of the trace instruction dispatch and execution, the trace data collection, route rebuild computation and traffic log inquiry.

Trace service console(s) executes two main operations: Once the trace request from the honeypot is received, it will mark the request with a serial number, then dispatches instructions to Trace agents(ITMs) in terms of the relevant traffic log; when the ITMs' feedback arrived, Trace service console would rebuild the attack route through the combined data and maintain a database for statistical inquiry that includes information as below: serial number, the request source, the initiation time, the agents' feedback data ( can act as proof), and provides user statistical inquiry.

Trace agents (ITMs) will analyze the incoming and outgoing traffic within a sliding time window once they captured the trace instructions. They detect the specific watermark feature using the relevant algorithm. The preliminary result would go back to the service console.

Trace service console has 4 modules: Time synchronization, Data synthesis processing, and Database and Communication control.Time synchronization provides unified time information for Trace service console and all of the Trace agents. Thus the trace events can be analyzed in proper association, and the database can use such given time for log.

Data synthesis processing combines all relevant information feedback from Trace agents (ITMs), and figures out an intrusion route rebuilt result. Data center, on the one hand, records all the relevant trace events, such as intrusion serial number, the trace request initiator, the feedback from all the agents (together with "proofing" data ), and the final route rebuild result; on the other hand, it enables user statistical inquiry with specific terms.

Communication control handles the initial trace request from the honeypot, marks a serial number for that request, dispatches instructions to Trace agents, and receives the trace feedback data.Trace agent includes Time synchronization, Traffic storage, Data analyze and Communication control. Time synchronization serves the same as in the console. Traffic storage provides the data center for analyze. Because the responding time (the time between the intrusion traffic pass the agent and the time agent get the trace instructions from console) is rather short due to automatic program, the agent will only reserve a "within sliding time window" traffic.

Data analyze pickups the source and destination IP address among the stored traffic, and verifies with the traffic logs to form a conclusion. Communication control module receives and feedbacks on the console's request.

### 6.1.3 Review of traffic flow

Honeypot has no service traffic with "outsider", so all network traffic connected to the honeypot can be regarded as be probed or intruded. Thus, by means of deploying some "forgery" sensitive data into the honeypot and monitoring any access to such data, we can precisely judge whether the honeypot is intruded or not.

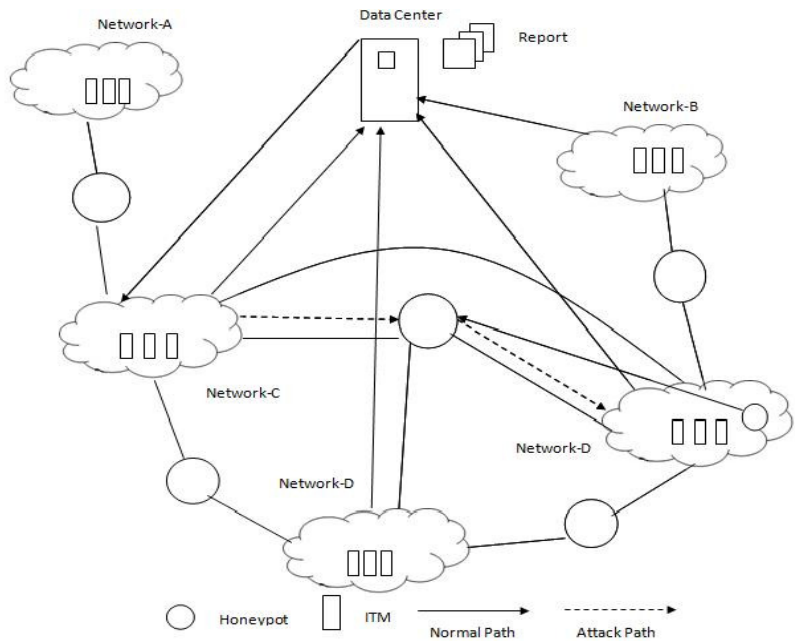

Fig.3 Tracbacking using Honeypot

Access to the entrap file with logs is monitored continually. Once such event occurred, the trace request will be initiated by the honeypot.

Once Trace service console (Data center) receives a trace request from honeypot, it will numbers it serially, then dispatch trace instructions to the agents in accordance to the specific traffic log information.

If the incoming traffic has not corresponding threshold, such incoming path is not the one we wanted. If the incoming check is positive while the outgoing negative, then the intrusion is within such connection. If the traffic goes to specific host, then such a host is the very troublemaker.

After all, if tracked threshold is found by the Trace agents, the relevant packets and the source/destination addresses will be reported to the Trace service console. According to all Trace agents' trace results, Trace service console generates the computation of the rebuilding of the final attack path and relevant log information such as: the request initiator, the event serial number, the initiation time, the trace result from Trace agents and relevant proofing data, and the final trace conclusion, and maps them to the database, After the path rebuilding, Trace service console may feedback the final result to the request initiator. Considering the honeypot is regarded as "captured", this feedback is somewhat unnecessary.

Compared to the existing trace back methods, this method has the following improvements:

1. The ITM can check and judge attack node or not by using the single packet instead of using more number of packets which is used in ICMP Traceback.

2. The efficiency is improved because of all processing here is fully automated.

3. Once the attack is detected, the traceback would go all out immediately, does not depends the status of the attack.

4. The number of packet markings is relatively less compared to the existing methods.

5. Scalability is high because, the process does not depend on the traffic volume or number of honeypots in the network.

6. Privacy is improved for the packets transferred over the network through the honeypots.

## 7. CONCLUSION AND FUTURE WORK

The approach integrates active real time flooding attack flow identification from botnet with determining required number of honeypots. The honeypot controller has been modelled at Data centre or ITMs to trigger honeypot generation in response to suspected attacks and route the attack traffic to honeypots. The performance of the proposed scheme is independent of attack traffic due to presence of honeypots at data centre or ITMs. It gives stable network functionality even in the presence of high attack load.

Some of the avenues for further extensions are with larger and heterogeneous networks. Back tracking can be applied on attack flows to reach the attack source. Both of them hold promise for evaluating and improving our DDoS detection and defence method and data centre information protection. The data centre load can be still minimized by used some distributed load sharing algorithms.